# Sense-Giving Strategies of Media Organisations in Social Media Disaster Communication:

# Findings from Hurricane Harvey

**Julian Marx**
Professional Communication in Electronic Media/Social Media
University of Duisburg-Essen
Duisburg, Germany
Email: julian.marx@uni-due.de

**Milad Mirbabaie**
Professional Communication in Electronic Media/Social Media
University of Duisburg-Essen
Duisburg, Germany
Email: milad.mirbabaie@uni-due.de

**Christian Ehnis**
Discipline of Business Information Systems
The University of Sydney
Sydney, Australia
Email: christian.ehnis@sydney.edu.au

## Abstract

Media organisations are essential communication stakeholders in social media disaster communication during extreme events. They perform gatekeeper and amplification roles which are crucial for collective sense-making processes. In that capacity, media organisations distribute information through social media, use it as a source of information, and share such information across different channels. Yet, little is known about the role of media organisations on social media as supposed sense-givers to effectively support the creation of mutual sense. This study investigates the communication strategies of media organisations in extreme events. A Twitter dataset consisting of 9,414,463 postings was collected during Hurricane Harvey in 2017. Social network and content analysis methods were applied to identify media communication approaches. Three different sense-giving strategies could be identified: retweeting of local in-house outlets; bound amplification of messages of individual to the organisation associated journalists; and open message amplification.

**Keywords** media organisations, social media, crisis communication, disaster communication, sense-giving

# 1   Introduction

Social media platforms are used as essential communication channels during crisis events. Different stakeholders such as individuals from the public, emergency management agencies (EMAs), non-government organisations (NGOs), and media organisations use social media to seek and share information. We know that individuals use these platforms to share their experiences or information (Imran et al. 2014), to organise help (Bunker et al. 2015), and for emotional support (Qu et al. 2011). EMAs use social media to provide trustworthy information, to share warnings, to engage with the public, or as a source (Reuter et al. 2012) for building situational awareness (Power and Kibell 2017). Literature in the domain of social media crisis communication mostly makes EMAs and their use of social media the focal point of research (e.g. Ahmed 2011; Palen et al. 2010), and investigates how EMAs are embedded in the collective sense-making process (Stieglitz et al. 2018; Weick 1988). Media organisations, too, have a crucial role in this process (Stieglitz et al. 2017) as they provide information and commentaries on the unfolding events. However, there has been little attention given to the role of journalistic coverage and its distribution in social media. At the same time, there is a knowledge gap on how information providers, as opposed to information seekers, contribute to collective sense-making. In an intra-organisational context, such actors are labelled as sense-givers (Giuliani 2016), but there is little known on how they contribute to public disaster coverage.

Disasters are chronicled from multiple sources by linking audio-visual media to social media messages, which enhances its perception. Particularly, users close to the site of events tend to include photographs and videos in their reporting (Bruns et al. 2012; Oh et al. 2010). This changes the nature of media coverage of disasters, as online communities can act as connected eyewitnesses who are not framed by mainstream media (Mortensen 2015). Traditional media outlets have entered the competition of social media news coverage and may apply the same logic to their practices. Media organisations may be considered to be sense-givers due to the fact that they pursue the interest of broadcasting information, thus, influencing perceptions on a disaster (Giuliani 2016; Pratt 2000). In this study, we focus on sense-giving strategies of media organisations in social media disaster communication. Therefore, the vital research questions are:

**RQ1**: How do media organisations distribute crisis information in social media?

**RQ2:** What are the strategies of media organisations to exert sense-giving during extreme events?

To uncover communication approaches of media organisations, we analyse their social media performance during a large scale natural disaster event – Hurricane Harvey in 2017. We created a Twitter dataset consisting of 9,414,463 tweets posted during the event. We apply social network analysis and content analysis techniques to interpret this data.

The paper is structured as follows. First, we provide an overview of the literature around crisis communication on Twitter and sense-making. Afterwards, we outline our research design, including data collection and data analysis procedures. Our findings section covers the outcomes of our data analysis. We then discuss the findings based on the theoretical construct. We conclude this paper with contributions, limitations, and implications for future research.

# 2   Literature Review – Crisis Communication on Twitter

In recent years, crisis communication participants have come to recognise that, Twitter, in particular, is the social medium of choice when it comes to exchanging crisis-related information online (Reuter et al. 2012). Disaster communications on Twitter includes the reporting of eyewitness accounts and relief activities (Hughes and Palen 2009; Oh et al. 2010). Twitter is not only a platform for sourcing information, but also soliciting donations, organising volunteers, publicising the names of missing persons, or broadcasting immediate needs during disaster events (Starbird and Palen 2012). While Twitter started out as a supplementary platform to link to external resources, the service has evolved to a source for mainstream media coverage itself (Bruns et al. 2012). Crawford (2010) describes Twitter as a network in which "rituals and transmissions are imbricated: communities of interest form clusters, and messages pass between them, with the occasional message being circulated to a much wider group" (p.150). The characteristics of Twitter, among other social media, added a novel logic to the dynamics between mass media, organisations and individuals (Dijck and Poell 2013). Literature often focusses on EMAs as organisations directly involved. EMAs may greatly benefit from voluntarily provided information on Twitter (Ehnis and Bunker 2013; Subba and Bui 2017). Studies indicate that EMAs often do not effectively position themselves as influential actors during crisis communication phases on Twitter, compared to other societal roles (Bakshy et al. 2011; Stieglitz et al. 2017). The validation and trustworthiness of information remains a crucial challenge (Schenk and Sicker 2011). Well-established media organisations and individual role types such as journalists, celebrities, or bystanders tend to amplify the dissemination of valuable information with greater magnitude (Mirbabaie and Zapatka 2017; Starbird and Palen 2012).

Influential communication participants can be characterised by large numbers of retweets of their posts, but also through actively retweeting others (Bruns et al. 2012). Local EMAs can often be regarded as topic authorities on Twitter communications which only contribute in a few discussions or relevant events (Pal and Counts 2011). In contrast, Media organisations tend to be general authorities, frequently publishing about a variety of topics and holding a large following. Scholars have examined credibility of journalistic coverage on Twitter (Castillo et al. 2011), and how disaster reporting influences traditional media coverage and vice versa (Valenzuela et al. 2017). Still, it is unclear how media reports diffuse through Twitter and gain decisive influence.

## 3 Theoretical Background

### 3.1 Deconstructing Sense-making through Role-based Approaches

Achieving a common understanding of the unfolding events during a disaster is important. Different communication participants vary in how much power and influence they can exert on inter-subjective knowledge creation. To uncover patterns, several studies employ a role-based approach by assigning social media users to different categories based on their characteristics (e.g. Blum et al. 2014; Ehnis et al. 2014). The study on hand might benefit from pre-defined classifications based on societal roles (Mirbabaie et al. 2014), network positioning (Mirbabaie and Zapatka 2017), and content (Blum et al. 2014). This allows researchers to better conceptualise how event participants utilise platforms to interact and communicate; carry out relationships and processes (Cornelissen 2012); and perform various tasks against the background of specific events.

Case study research of this kind aims to reproduce relational patterns between involved entities (Kohlbacher 2006). Those are still somehow attached to the events of the corresponding crisis case, but might be transferred onto other contexts if bundled under the superstructure of role-specific archetypes of social media communication. Those types may be determined through breaking down social expectations and integrate users into a spectrum of role clusters, i.e. political groups, individuals, commercial organisations, non-governmental organisations, EMA, and media organisations (Mirbabaie et al. 2014).

### 3.2 Making and Giving Sense in Crisis Situations

We employ sense-giving as a subsumable concept of sense-making. Originating in the context of organisational theory, sense-making was extensively devised by Karl Weick, who qualifies it as a process that "*unfolds as a sequence in which people [...] engage ongoing circumstances from which they extract cues and make plausible sense retrospectively, while enacting more or less order into those ongoing circumstances*" (Weick et al. 2005, p.409). Crisis situations, particularly, surface the need for sense-making within organisations due to low probability of events and involved actors being unprepared (Weick 1988). Scholars have shown eclectic creativity on how to transfer the concept of sense-making onto other domains and contexts. The applicability of sense-making within the sphere of public crisis communication seems indisputable. Inverting the observations that could be made within organisations to an unbound societal perspective holds promising realisations for research on public crisis communication. Communities may collectively work toward mutual sense by means of social media, or find even more uncertainty on an individual level if collective efforts fail (Dailey and Starbird 2015).

The contra effect of information seeking and its provision is mirrored in sense-making theory. Sense-demanding (Vlaar et al. 2008) is up against sense-giving (Pratt 2000), or even sense-breaking (Giuliani 2016). While these sub-concepts describe possible occurrences or deliberate actions during the sense-making process, they are not exclusive. A single communication participant may exert effects on more than one stream of meaning, e.g. performing sense-demanding and sense-giving efforts on different fronts of knowledge. Sense-giving describes the attempt to influence or change the way others perceive a situation and steer actions towards a favourable direction (Aanestad and Blegind 2016). The way crisis communication participants perform sense-giving is yet to be investigated. Therefore, our study aims to explore acts of sense-giving by media organisations during Hurricane Harvey.

## 4   Case Background – Hurricane Harvey 2017

In late August of 2017, the coastline of Texas and parts of Louisiana faced the destructive force of category 4 hurricane 'Harvey'. It approached the mainland of Texas near Corpus Christi on the 26th of August. Officials declared a state of disaster and several counties decided to evacuate residential areas. The storm resulted in severe floods and winds had gust speeds up to 215 km/h. Shortly after, Harvey reached the urban areas of Houston and continued to release heavy rainfalls that swamped large parts of the city. The hurricane lost strength on August 30th after which recovery and reconstruction measurements commenced (Sternitzky-DiNapoli 2017). In contrast to human-induced incidents such as terror attacks or other unforeseen events, Hurricane Harvey had the character of a predictable crisis. Weather forecasts gave insight for planning and anticipated the effects of the event before unfolding. Strong public interest and extensive media coverage developed in the course of the crisis.

## 5   Research Design

### 5.1   Data Collection

The projected case study builds upon a dataset of relevant Twitter data concerning Hurricane Harvey. By the means of a self-developed Java crawler and the open source library Twitter4J3, six days of Twitter communication from the 26th (0:00 UTC) to the 31st (23:59 UTC) of August 2017 was collected. To obtain only relevant data, only content containing the keywords *hurricaneharvey, harvey, hurricane* was crawled. The selection of keywords was based on their usage frequency during the crisis as well as the predominant usage of hashtags, which are included in the selected keywords. Moreover, the crawler was set to only gather data provided with English language settings. This data tracking resulted in a dataset of 9,414,463 tweets. The extracted data was stored in a MySQL database for further pre-processing.

### 5.2   Data Analysis

The data was analysed with social network analysis and content analysis techniques (Stieglitz et al. 2018a). To utilise these research methods, the dataset needed to be processed further. First, the dataset needed to match the requirements of a social network analysis tool. Second, the data material had to be made accessible for content analysis procedures. To determine influence on Twitter, we turn to literature that suggests a combined set of metrics including original tweets, retweets, mentions as well as characteristics from graph theory (Pal and Counts 2011). Subsequent to methodical steps involving data analytics, collating a study's findings with network and content behavioural archetypes from the literature might be substantial towards understanding the sense-making efforts of single user roles (Mirbabaie and Zapatka 2017; Blum et al. 2014). Hence, the research design of this study is constructed upon a combined approach of network metrics and user roles. To increase the manageability of the dataset, it was divided into six segments consisting of tweets for each day. The number of segments was determined to match the criteria of (a) complying with the chronological sequence of events and (b) conforming to the requirements of analysis tools, i.e. reducing the data volume for each social network analysis (Stieglitz et al. 2018b).

#### 5.2.1   Social Network Analysis

Data was processed in order to be displayed in a social network analysis tool, including the configuration of both edges (links) and nodes (vertices) of the network (graph). According to graph theory, centrality measures of may reveal characteristics and social ties of certain network participants (Wasserman and Faust 1994). In a directed retweet network, the in-degree represents the number of retweets a user received and serves as a strong indicator for the reach of a message (Kwak et al. 2010). Power users are defined as the communication participants who receive the highest number of retweets in relation to the frequency of their overall output (Oh et al. 2015). We use these power users as an approximation for influential users. To obtain representative power users of predefined communication roles, accounts were ranked based on their in-degree value and subsequently assigned to the fitting role category. This sample of power users provided a manageable set of tweets for qualitative content analysis. We used social network analysis procedures to (1) compile a sample of influential tweets, and (2) reveal relevant power users, in particular media organisations.

#### 5.2.2   Content Analysis

To determine what types of messages were influential during Hurricane Harvey, we created a sample of power users and their respective tweet sets. We focused on power users, because each media organisation is classified as a power user. By implication they hold influential positions in the network and, therefore, complies with our definition of sense-givers. We scanned through the daily ranking of top 100 power users in descending order until each role type (adapted from Mirbabaie and Zapatka 2017) occurred at least five times among the most visible power users. The six daily rankings per role were merged to one overall ranking to eliminate double entries. This was necessary to ensure an uncluttered view on the entire activity profile of potentially influential media-related accounts over the course of the crisis. We then coded the tweets of journalists and media organisations on the basis of a self-developed codebook, following the guidelines of Mayring (2000, 2014). It was developed using pre-defined role types from the literature (Mirbabaie 2014; Stieglitz et al. 2017), to distinguish media organisations and journalists from each other and remaining communication roles. After separately coding the sample, the inter-rater reliability was calculated to verify if all coders used the codebook in an equal manner. Using Krippedorff's alpha, a score of .887 was calculated. Our coding can be rated as reliable as α ≥ .800 (Krippendorff 2013).

## 6　Findings

In total, there were 9,414,463 tweets in the dataset, authored by 1,093,349 unique users. Considering our power-user sample of 100 influential accounts per day, media organisations made up the most active role type during the six-day period, as 57.7% of the sample's content stems from involved media outlets, followed by EMA (15.5%) and journalists (8,8%). These figures, as shown in table 1, relate only to original tweet frequency, with no consideration of the actual reach of such messages.

| Role type | Number of accounts | Tweet volume | Percentage |
| --- | --- | --- | --- |
| Media Organisation | 12 | 734 | 57.70% |
| EMA | 12 | 198 | 15.50% |
| Journalists | 20 | 112 | 8.80% |
| Private Person | 20 | 69 | 5.40% |
| Politicians | 18 | 56 | 4.40% |
| Influencers/Bloggers | 20 | 55 | 4.30% |
| Celebrities | 19 | 50 | 3.90% |

*Table 1. Activity of top 5 power users per role type and per day (combined).*

In order to serve our research objective, we narrow down the subsequent analysis to media organisations and journalists. The resulting sub-sample contains a total of 846 tweets. As part of the content analysis, we classified each tweet according to its information type. The underlying codebook was equipped with 17 different information types. As with the role classification, only occurring information types were listed in table 2. This listing also includes retweeting or replying one of the information types, e.g. a journalist retweeting an official statement.

| Role type (Tweets) | Official statement | News/Crisis Information | Personal opinion | Personal Experience | Forwarding message | Solicitousness | Marketing |
| --- | --- | --- | --- | --- | --- | --- | --- |
| Media Organisation (739) | 11(1%) | 626(85%) | 0(0%) | 40(5%) | 54(8%) | 3(1%) | 5(1%) |
| Journalist (73) | 0(0%) | 66(60%) | 12(11%) | 11(10%) | 18(16%) | 3(3%) | 0(0%) |

*Table 2. Information types used by each role*

Media organisations primarily focussed on the distribution of crisis information and news (85%). The numbers of crisis information provided by media organisations have the highest absolute score in the sample (626). Journalists' postings, too, contain mostly crisis information (60%).

To take another glimpse beneath the surface of how information is being disseminated through media organisations, illustrative examples of this role type and an in-depth analysis of representing accounts provides valuable insight. The initial codebook provided a distinction between national media outlets and international media organisations. However, throughout the analysis, an additional subcategory emerged within the realm of media organisations. Whereas international media organisations are not represented in our sample, a clear distinction could be made between national and regional media organisations. The sample contained a total of 12 accounts that represent a media outlet, which are listed in table 3. Among those accounts, one account (@thehill), received retweets from postings prior to the tracking, but was not authoring about Hurricane Harvey during the six-day period. Nine accounts actively tweeted and are considered national (or widely recognised) media outlets (@MSNBC, @NBCNews, @washingtonpost, @ABCNews, @CBSNews, @CNN, @FoxNews, @AP, @nytimes), and two are classified as regional media outlets (@abc13houston, @HoustonChron).

| Media outlet | Tweets | Retweets | Replies | T/h* | Indegree** | Betweenness centrality*** |
|---|---|---|---|---|---|---|
| ABC News | 4 | 1 | 0 | 0.03 | 30,310(2) | 0.00 |
| ABC13Houston | 73 | 65 | 0 | 0.96 | 12,768(5) | 120,533,523.78 |
| CBS News | 4 | 3 | 0 | 0.05 | 13,705(2) | 480,742.74 |
| CNN | 22 | 2 | 0 | 0.17 | 44,828(4) | 16,535,160.94 |
| Fox News | 125 | 0 | 0 | 0.87 | 17,419 (6) | 484,511.93 |
| Houston Chronicle | 266 | 99 | 1 | 2.54 | 8,618(4) | 110,538,287.61 |
| MSNBC | 0 | 7 | 0 | 0.05 | 17,516(1) | 7952881.12 |
| NBC News | 33 | 11 | 0 | 0.31 | 38,472(2) | 170,269,093.88 |
| New York Times | 0 | 2 | 0 | 0.01 | 7,973(6) | 26,062,251.89 |
| The Associated Press | 3 | 12 | 0 | 0.1 | 13,137(6) | 485,304.48 |
| The Hill | 0 | 0 | 0 | 0 | 19,123(1) | 0.00 |
| Washington Post | 1 | 0 | 0 | <0.01 | 12,530(1) | 230,110.90 |

*tweet frequency=original tweets+retweets+replies/6*24 hours
**highest daily indegree value during 6-day period (day)
***value corresponding to max. indegree (same day)

*Table 3. Media organisation's total metrics*

Some of the national media organisations focussed on retweeting information from external sources (@MSNBC, @AP, @nytimes). Others produced content and forwent retweeting (@washingtonpost, @CNN, @FoxNews). The regional outlets, however, show significantly higher numbers in terms of creating original content as well as retweeting other sources. The Houston Chronicle Twitter account, which represents a local daily newspaper, sent out 2.54 tweets per hour, whereas the most frequently tweeting national media outlet (@FoxNews) only posted 0.87 Harvey-related tweets per hour. At the same time, both regional media outlets score significantly high on betweenness centrality. Frequently authoring and retweeting relevant tweets, a high indegree and betweenness centrality are indicators for highly influential users (Pal and Counts, 2011). Table 4 shows the different approaches of both media outlets on how to distribute information. The @HoustonChron account retweeted almost exclusively journalists. The three exceptions occurred, when the Houston Chronicle account retweeted local EMA.

| Regional media outlet | Performed retweets | Retweeted private persons | Retweeted journalists | Retweeted others |
|---|---|---|---|---|
| ABC 13 Houston | 65 | 12 | 28 | 25 |
| Houston Chronicle | 99 | 0 | 96 | 3 |

*Table 4. Comparison of regional media outlets' retweet activity*

Taking a closer look at the retweeted journalist accounts, revealed that every single account (28 unique users) belongs to reporters/employees of the Houston Chronicle. The account frequently retweeted eyewitness reports experienced by their reporters (e.g. Chron_MattYoung). After retweeting this eyewitness report, which was complemented by an image, the @HoustonChron account turned the information into an original tweet.

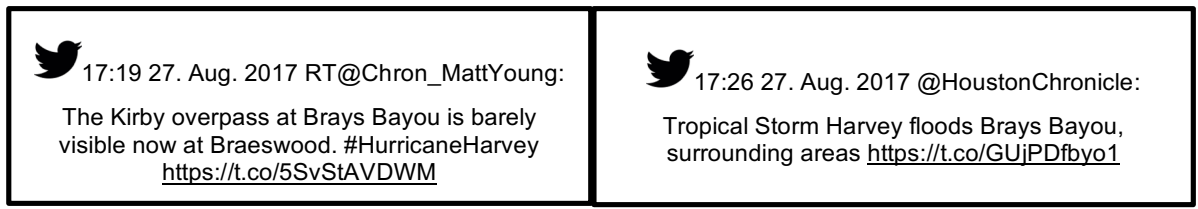

Other than @HoustonChron, the Twitter presence of the regional TV station @abc13houston included eyewitness reports authored by private persons into their publishing strategy. In an earlier tweet, @abc13houston encouraged their followers to submit eyewitness reports using #ABC13Eyewitness.

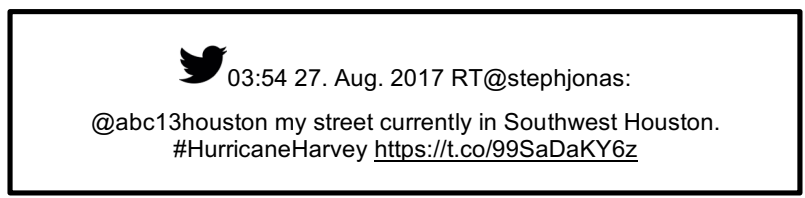

The journalists retweeted by @abc13houston were, similar to @HoustonChron, reporters of their own branch. Figure 1 shows the ego-network graphics of both media outlets on August 26th, the first day of the six-day period. It visualises all retweets that were sent from each account (@abc13houston's outdegree=33; @HoustonChron's out-degree=22). The size of a node indicates the influence of an account on that day (indegree value). The ten users with the highest indegree were labelled for more clarity. The colour of an edge as well as the node colours emphasise the role type of an account (blue=media organisation, green=EMA, grey=private person, red=politician, purple=journalist).

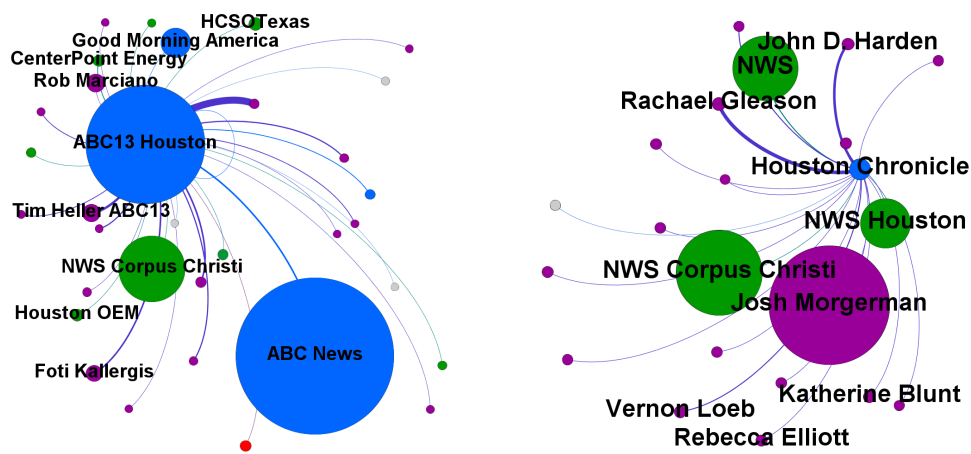

*Figure 1: Ego-network @abc13houston (left) and ego-network @HoustonChron (right)*

Both local media outlets differ in how influential their accounts were with regards to the indegree. @abc13houston was retweeted 8748 times, whereas @HoustonChron counts an indegree value of 652. Both accounts retweeted mostly journalists. The @abc13houston account forwarded messages of 18 journalists, 1 politician, 3 private persons, 8 EMAs and 3 other media organisations. The @HoustonChron account, however, limited their retweet recipients to 3 EMA, and 19 journalists, who are officially denoted as reporters of the Houston Chronicle. Another observation pointing to the dynamics of national, regional and on the spot information can be made by looking at the retweet behaviour of @AP, which is the Twitter account of the nationwide media outlet "The Associated Press". The activity of this account amounts for retweeting @APCentralRegion 10 times (out of 12 total retweets). In this case, a nationwide media outlet retweeted its regional branch. The @abc13houston account, in contrast, did not receive any retweets by @ABCNews.

## 7 Discussion

Due to the predictable character of the Hurricane Harvey crisis, official institutions as well as media organisations, were able to pre-arrange social media communication strategies and put systems of coverage in place. Our data shows that media organisations author a large number of unique content, but also function as bridges between users of other role categories. The examination of media organisations as a central communication role during Hurricane Harvey revealed three strategies of media organisations to leverage their social media reach during crisis events. As shown in figure 2, media organisations can use (1) popularity arbitrage by retweeting their local in-house outlet. During Hurricane Harvey, The Associated Press (@AP) did rarely author original content related to the crisis. Instead, their nation-wide account focussed on retweeting @APCentralRegion.

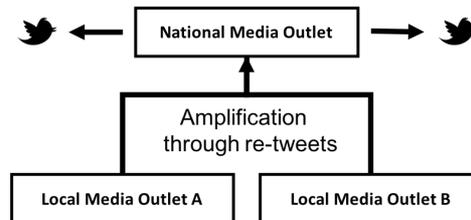

*Figure 2: Strategy 1- Popularity arbitrage by retweeting their local in-house outlets*

Another strategy to increase network-related influence is (2) bound amplification through quality assured sources such as in-house journalists. The local newspaper @HoustonChron strategically retweeted journalists exclusively reporting for the Houston Chronicle. Due to the large number of individual journalists (28), this accumulation of eyewitness reports covered a major part of the crisis region as the reporters were distributed over the entire Houston area. It scaled the reach of those messages as all 28 journalists used their individual Twitter account to instantaneously report from the crisis scene. Eventually, their content was edited and uploaded to the Houston Chronicle website and posted as an original tweet via @HoustonChron. At the same time, this strategy (figure 3) served as quality assurance – a factor of trustworthiness personal experiences and eyewitness reports often do not provide (Schenk and Sicker 2011).

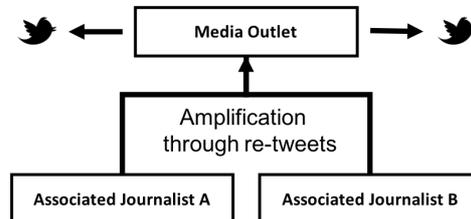

*Figure 3: Strategy 2- Bound amplification through quality assured sources such as in-house journalists*

Including public tweets in an organisation's Twitter activity leads to (3) open amplification. Demonstrated by @abc13houston, this strategy involves eyewitness reports of any role type but primarily private persons. During Hurricane Harvey, the news outlet even called out for contributions by using the hashtag #ABC13Eyewitness or mentioning @abc13houston in the tweet. This procedure does not only engage members of the public to actively report crisis information but also serves as a mechanism to overcome information overload. Even though @abc13houston performed a well-organised social media strategy and positioned themselves as an authority in terms of both network and content, potential leverage was left unattended. The popular parent account @ABCNews refrained from retweeting its regional branch. Scholars defined the issue of effectively listening and filtering social media crisis communication as one of the most crucial challenges for crisis responders (Oh et al., 2010). Leveraging influence as shown in figure 4 clearly addresses this shortcoming among crisis responders and might be useful for practical implications.

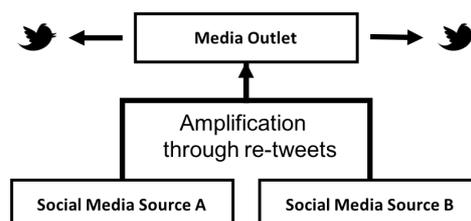

*Figure 4: Strategy 3- Open amplification of social media sources*

Overall, media organisations were found to leverage their influence and, therefore, contribute to sense-making by providing crisis information. This qualifies media organisations to be seen as sense-givers who support collective knowledge creation (Aanestad and Blegind 2016). Sense-giving is an attempt to steer actions towards a favoured direction (Giuliani 2016). Media organisations who focalise sense-giving efforts are capable of creating an information prevalence. These efforts, if information are accurate, may result in improved knowledge or relief actions. In our case, local media outlets such as @HoustonChron and @abc13houston gathered public information through retweeting reporters and/or private persons. This three-level hierarchy of national, local, and individual journalism comprises of useful strategies to reduce complexity and benefit collective sense-making.

## 8  Conclusion and Further Research

This study attempts to unfold social media communication strategies of media organisations during disasters. To that matter, we conducted a content analysis on Twitter postings that were distributed during the critical phase of Hurricane Harvey in 2017. We combined this method with social network analysis procedures to be able to carry out a more precise examination of the media outlets that were most engaged in the communication network. Our results reveal local media outlets, in particular, play a critical role in disseminating of crisis information, and thus, to be decisive sense-givers. We identified three different sense-giving strategies of media organisations: retweeting local in-house outlets; bound amplification of messages of individual associated journalists; and open message amplification.

Our study contributes to the existing body of knowledge on social media disaster communication as it emphasises the standing of media organisations as an authority to complement crisis response operations. Whereas most scholars focus on enhancing social media utilisation for EMAs, we bring attention to media organisations. This provides a more differentiated view on the coexistence of certain societal roles which hold stakes in social media crisis communication. Additionally, we pursued to translate the notion of sense-giving onto the domain of social media disaster communication. Theorising our findings in this way may help to reach an improved understanding of how knowledge is being created during extreme events. EMAs should have the conception of themselves to be a publishers of crisis information during extreme events. EMAs, too, are often organised as local and nationwide branches. Hence, the same amplification mechanics concern EMAs, which may take publishing strategies of media organisations as a paragon for information distribution.  Our study comes with limitations as we focused on Twitter data exclusively, which has specific characteristics. This snippet of communication recording does not cover the entire crisis communications, but mirrors social media's major share, as Twitter has become leading-edge for disaster coverage. Moreover, restricting the data collection to certain keywords and a confined timespan may exclude relevant messages. Even though the six-day period covers a majority of the Harvey-related Twitter communication, our tracking missed postings from before and after the main events.

Further research in this area might cover the challenges for media organisations when facing an unpredictable crisis event. Delving deeper into this could encompass interviews with media organisations. Moreover, one could extend the investigation of crisis coverage to other social media platforms and examine the intersection of social media strategies with traditional media reporting. On a theoretical level, adjoining sub-concepts of sense-making such as sense-demanding or sense-breaking with regards to media organisations seem worthwhile to investigate.

## 9  References

Aanestad, M., and Blegind, T. 2016. "Collective Mindfulness in Post-Implementation IS Adaptation Processes" in *Information and Organization* (26:1), pp. 13–27.

Ahmed, A. 2011. "Using Social Media in Disaster Management" in *Proceedings of the International Conference on Information Systems,* Shanghai: China, pp. 16–27.

Bakshy, E., Hofman, J. M., Mason, W. A., and Watts, D. J. 2011. "Everyone's an Influencer" in *Proceedings of the ACM International Conference on Web Search and Data Mining*, Hong Kong: China, pp. 65–74.

Blum, J., Kefalidou, G., Houghton, R., Flintham, M., Arunachalam, U., and Goulden, M. 2014. "Majority Report: Citizen Empowerment through Collaborative Sensemaking" in *Proceedings of the International Conference on Information Systems for Crisis Response and Management*, University Park, Pennsylvania: USA, pp. 767–771.

Bruns, A., Burgess, J., Crawford, K., and Shaw, F. 2012. "#qldfloods and @QPSMedia: Crisis

Communication on Twitter in the 2011 South East Queensland Floods" in *Brisbane: ARC Centre of Excellence for Creative Industries and Innovation*, pp. 1–57.

Bunker, D., Sleigh, T., Levine, L., and Ehnis, C. 2015. "Disaster Management: Building Resilient Systems to Aid Recovery" in *Research Proceedings from the Bushfire and Natural Hazards CRC & AFAC Conference*, pp. 1–6.

Castillo, C., Mendoza, M., and Poblete, B. 2011. "Information Credibility on Twitter" in *Proceedings of the International Conference on World Wide Web,* Hyderabad: India, pp. 675-684.

Cornelissen, J. P. 2012. "Sensemaking Under Pressure: The Influence of Professional Roles and Social Accountability on the Creation of Sense" in *Organization Science* (23:1), pp. 118–137.

Crawford, K. 2010. "News To Me: Twitter and the Personal Networking of News" in *News Online*, pp. 135–156.

Dailey, D., and Starbird, K. 2015. ""It's Raining Dispersants" Collective Sensemaking of Complex Information in Crisis Contexts" in *Proceedings of the ACM Conference on Computer Supported Cooperative Work & Social Computing*, Vancouver, BC: Canada, pp. 155–158.

Dijck, J. Van, and Poell, T. 2013. "Understanding Social Media Logic" in *Media and Communication* (1:1), pp. 2–14.

Ehnis, C., and Bunker, D. 2013. "The Impact of Disaster Typology on Social Media Use by Emergency Services Agencies: The Case of the Boston Marathon Bombing" in *Proceedings of the Australasian Conference on Information Systems,* Melbourne: Australia, pp. 1–12.

Ehnis, C., Mirbabaie, M., Bunker, D., and Stieglitz, S. 2014. "The Role of Social Media Network Participants in Extreme Events" in *Proceedings of the Australian Conference of Information Systems*, Auckland: New Zealand, pp. 1–10.

Giuliani, M. 2016. "Sensemaking, Sensegiving and Sensebreaking" in *Journal of Intellectual Capital* (17:2), pp. 218–237.

Hughes, A., and Palen, L. 2009. "Twitter Adoption and Use in Mass Convergence and Emergency Events" in *International Journal of Emergency Management* (6:3/4), p. 248-260.

Imran, M., Castillo, C., Diaz, F., and Vieweg, S. 2014. "Processing Social Media Messages in Mass Emergency: A Survey" in *ACM Computing Surveys* (47:4), pp. 1-38.

Kohlbacher, F. 2006. "The Use of Qualitative Content Analysis in Case Study Research" in *Forum: Qualitative Social Research* (7:1), pp. 1–30.

Krippendorff, K. 2013. *Content Analysis: An Introduction to Its Methodology*, (3rd edition), Thousand Oaks, CA: Sage.

Kwak, H., Lee, C., Park, H., and Moon, S. 2010. "What Is Twitter, a Social Network or a News Media?" in *Proceedings of the International World Wide Web Conference*, Raleigh, North Carolina: USA, pp. 1–10.

Mayring, P. 2000. "Qualitative Content Analysis," in *Forum Qualitative Research* (1:2), p. 1-10.

Mayring, P. 2014. *Qualitative Content Analysis: Theoretical Foundation, Basic Procedures and Software Solution*, Klagenfurt.

Mirbabaie, M., Ehnis, C., Stieglitz, S., and Bunker D. 2014. "Communication Roles in Public Events - A Case Study on Twitter Communication" in *Proceedings of the Working Conference on Information Systems and Organizations*, Auckland: New Zealand, pp. 207–218.

Mirbabaie, M., and Zapatka, E. 2017. "Sensemaking in Social Media Crisis Communication - A Case Study on the Brussels Bombings in 2016" in *Proceedings of the European Conference on Information Systems*, Guimarães: Portugal, pp. 2169–2186.

Mortensen, M. 2015. "Connective Witnessing: Reconfiguring the Relationship between the Individual and the Collective," in *Information Communication and Society* (18:11), pp. 1393–1406.

Oh, O., Eom, C., and Rao, H. R. 2015. "Role of Social Media in Social Change: An Analysis of Collective Sense Making During the 2011 Egypt Revolution" in *Information Systems Research* (26:1), pp. 1–14.

Oh, O., Kwon, K. H., and Rao, R. H. 2010. "An Exploration of Social Media in Extreme Events: Rumor

   Theory and Twitter during the Haiti Earthquake 2010" in *Proceedings of the International Conference on Information Systems,* St. Louis: USA, pp. 1–15.

Pal, A., and Counts, S. 2011. "Identifying Topical Authorities in Microblogs" in *Proceedings of the ACM International Conference on Web Search and Data Mining*, Hong Kong: China, p. 45-54.

Palen, L., Anderson, K. M., Mark, G., Martin, J., Sicker, D., Palmer, M., and Grunwald, D. 2010. "A Vision for Technology-Mediated Support for Public Participation & Assistance in Mass Emergencies & Disasters" in *Proceedings of ACM-BCS Visions of Computer Science*, Edinburgh: United Kingdom, pp. 1–12.

Power, R., and Kibell, J. 2017. "The Social Media Intelligence Analyst for Emergency Management" in *Proceedings of the Hawaii International Conference on System Sciences*, pp. 313–322.

Pratt, M. G. 2000. "The Good, the Bad, and the Ambivalent: Managing Identification among Amway Distributors," in *Administrative Science Quarterly* (45:3), p. 456-493.

Qu, Y., Huang, C., Zhang, P., and Zhang, J. 2011. "Microblogging after a Major Disaster in China" in *Proceedings of the ACM Conference on Computer Supported Cooperative Work*, Hangzhou: China, p. 25-34.

Reuter, C., Marx, A., and Pipek, V. 2012. "Crisis Management 2.0" in *International Journal of Information Systems for Crisis Response and Management* (4:1), pp. 1–16.

Schenk, C. B., and Sicker, D. C. 2011. "Finding Event-Specific Influencers in Dynamic Social Networks" in *Proceedings of the IEEE International Conference on Privacy, Security, Risk and Trust*, Boston: USA, pp. 501–504.

Starbird, K., and Palen, L. 2012. "(How) Will the Revolution Be Retweeted?: Information Diffusion and the 2011 Egyptian Uprising" in *Proceedings of the ACM Conference on Computer Supported Cooperative Work*, pp. 7–16.

Sternitzky-DiNapoli, D. 2017. Timeline of Hurricane Harvey, for those who don't know what day it is, *Houston Chronicle*. Last retrieved on October 10th 2018: https://www.chron.com/news/houston-weather/hurricaneharvey/article/Hurricane-Harvey-timeline-12169265.php

Stieglitz, S., Meske, C., Ross, B., and Mirbabaie, M. 2018. "Going Back in Time to Predict the Future - The Complex Role of the Data Collection Period in Social Media Analytics" in *Information Systems Frontiers* (June), pp. 1–15.

Stieglitz, S., Mirbabaie, M., Fromm, J., and Melzer, S. 2018. "The Adoption of Social Media Analytics for Crisis Management - Challenges and Opportunities" in *Proceedings of the European Conference on Information Systems,* Portsmouth: United Kingdom, pp. 1-19.

Stieglitz, S., Mirbabaie, M., and Milde, M. 2018. "Social Positions and Collective Sense-Making in Crisis Communication" in *International Journal of Human-Computer Interaction* (34:4), pp. 328–355.

Stieglitz, S., Mirbabaie, M., Schwenner, L., Marx, J., Lehr, J., and Brünker, F. 2017. "Sensemaking and Communication Roles in Social Media Crisis Communication" *Proceedings of the Internationale Tagung Wirtschaftsinformatik,* St. Gallen: Switzerland, pp. 1333–1347.

Subba, R., and Bui, T. 2017. "Online Convergence Behavior, Social Media Communications and Crisis Response: An Empirical Study of the 2015 Nepal Earthquake Police Twitter Project" in *Proceedings of the Hawaii International Conference on System Sciences*, pp. 284–293.

Valenzuela, S., Puente, S., and Flores, P. M. 2017. "Comparing Disaster News on Twitter and Television: An Intermedia Agenda Setting Perspective" in *Journal of Broadcasting and Electronic Media* (61:4), pp. 615–637.

Vlaar, P., van Fenema, P., and Tiwari, V. 2008. "Cocreating Understanding and Value in Distributed Work" in *MIS Quarterly* (32:2), pp. 227–255.

Wasserman, S., and Faust, K. 1994. *Social Network Analysis Methods And Applications*, Cambridge University Press.

Weick, K. E. 1988. "Enacted Sensemaking in Crisis Situations," in *Journal of Management Studies* (25:4), pp. 305–317.

Weick, K. E., Sutcliffe, K. M., and Obstfeld, D. 2005. "Organizing and the Process of Sensemaking" in *Organization Science* (16:4), pp. 409–421.